\documentclass{article}
\usepackage{amsfonts}
\usepackage{graphicx}
\usepackage{graphics}
\newcommand{\R}{\mathbb{R}}

\begin{document}

\title{Ontology from Local Hierarchical Structure in Text}

\author{F. Murtagh, Royal Holloway, University of London \\
Egham TW20 0EX, UK  \\
{\tt fionn@cs.rhul.ac.uk}  \\
 \\
J. Mothe, Institut de Recherche en Informatique de Toulouse \\ 
118, Route de Narbonne, 31062 Toulouse Cedex 04, France \\
{\tt mothe@irit.fr} \\
 \\
and \\
 \\
K. Englmeier, Fachhochschule Schmalkalden \\
Blechhammer, 98574 Schmalkalden, Germany \\
{\tt k.englmeier@fh-sm.de} }

\maketitle

\begin{abstract}
We study the notion of hierarchy in the context of visualizing 
textual data and navigating text collections.  A formal framework for 
``hierarchy'' is given by an ultrametric topology.  This provides us
with a  theoretical foundation for concept hierarchy creation. 
A major objective is {\em scalable} annotation or labeling of concept maps.
Serendipitously we pursue other objectives such as  
deriving common word pair (and triplet) phrases, i.e., 
word 2- and 3-grams.  
We evaluate our approach using (i) a collection of texts, 
(ii) a single text subdivided into successive parts (for which we provide an 
interactive demonstrator), and (iii) a text
subdivided at the sentence or line level.  While detailing a generic 
framework, a distinguishing feature of our work is that we focus on 
{\em locality} of hierarchic structure in order to extract semantic 
information.  
\end{abstract}

\bigskip

\noindent
{\bf Categories and Subject Descriptors:}
H.5 (Information interfaces and presentation), I.5.3 (Clustering),
H.5.2 (User interfaces), I.7.2 (Document preparation), H.3 (Information
storage and retrieval)

\section{Introduction}

Since the mid-1990s we built visual interactive maps of bibliographic and
database information at Strasbourg Astronomical Observatory, and some of
these, with references, are available at Murtagh [2006d].  The 
automated annotation of such maps is not easy.  
At the time of writing\footnote{Zdnet:
http://news.zdnet.co.uk/itmanagement/0,1000000308,39284764,00.htm  BBC:
http://www.bbc.co.uk/radio4/history/inourtime/inourtime.shtml}
 Zdnet and the BBC 
(British Broadcasting Corporation) use interactive annotated maps to 
support information navigation.  In Zdnet's 
case, some prominent terms are graphically presented and can be used to 
carry out a local search; and in the BBC case, terms relating to 
downloadable radio programs are displayed in moving sizes and locations.  

In the work described in this article, 
we adopt a different approach: we select the terms of interest
in a manual or semi-automated way.  This not only 
represents expert user judgement but also allows for inclusion of
rare or very frequent terms.  In one of our three case studies,
we use an automated way to select such terms. 
%Above and beyond the annotation of information maps we describe the 
%theory and how we can automate most other elements of such approaches to
%information structuring and display, and in their  use of
%information navigation.
For selected terms, we use their inter-relationships to build a hierarchy
and use this as a central device 
for summarizing information and supporting navigation.

``Ontologies are
often equated with taxonomic hierarchies of classes ... but ontologies
need not be limited to such a form'' [Gruber 2001].  Gruber is 
cited in G\'omez-P\'erez et al.\ [2004] as characterizing an ontology as 
``an explicit specification of a conceptualization''.  In Wache et al.\ 
[2001], ontologies are motivated by semantic heterogeneity of distributed
data stores.  This is also termed data heterogeneity and is counterposed 
to structural or schematic heterogeneity.  Ontologies are motivated 
by Wache et al.\ [2001] ``for the explication of implicit and hidden 
knowledge'', as ``a possible approach to overcome the problem of 
semantic heterogeneity''.  So, ontologies may help with integration of 
diverse, but related, data; or they may help with clarifying or 
disambiguating distinctions in the heterogeneous data.  Ontologies are likely 
to be of immediate help in supporting querying.  For example, the query 
model may be based on the ontology (or ontologies) used.  
% Ontology representations; and mappings between ontologies. 

There is extensive activity on standards and software, relating 
more to  the above-mentioned schematic 
rather than semantic heterogeneity, and a useful survey of this area 
is Denny [2004].  Denny takes an ontology in a broad-ranging view
 as a knowledge-representation scheme.  

\subsection{Short Review of Methods Used}

A short review of some recent approaches in this area follows.  
Ahmad and Gillam 
[2005] develop a semi-automated approach using text with no markup.  Multiword
expressions are determined, and frequency of occurrence information is 
used to point to term or phrase  importance. A stop list is used to avoid 
irrelevant words.  Part of speech analysis is not used.
 A semantic net is formed to allow development of the ontology elements.  

Abou Assali and Zanghi [2006] use syntactic part of speech tagging 
to determine the nouns.  
These authors retain sufficiently frequent nouns.  They apply the notion of 
weak subsumption: if -- for the most part -- 
 a word is in a text that another is in,
and not vice versa, then this leads to a hierarchical relationship.

Chuang and Chien [2005] assert that multiway trees are appropriate for 
concept hierarchies, whereas binary trees are built using hierarchical 
clustering algorithms.  Hence they modify the latter to provide more 
appropriate output.  (A formal approach for mapping a binary hierarchical 
classification tree onto a multiway hierarchy is described in Murtagh [2006b].)

A hierarchical clustering has often been used to represent an ontology.  
Note that this is usually not a concept hierarchy.  A concept hierarchy 
is based on a subsumption relationship between terms, whereas a hierarchical
clustering is an embedded set of clusters of the term set.  Later in this
article (section \ref{secorientree}), we show a way to derive 
a concept hierarchy, involving subsumption of terms, 
from a hierarchic clustering.  

A hierarchic clustering is typically a binary, rooted, terminal 
labeled, ranked tree, and a concept hierarchy is typically 
a multiway, rooted, terminal and non-terminal labeled, ranked tree.  
By starting with the former (binary) tree representation, we have an
extensive theoretical and formal arsenal at our disposal, to represent 
the main lines of what we need to do, and to help to avoid ad hoc, 
user parameter-based, ``engineering'' approaches.  As seen later in this
work, we start by laying the foundations of our perspective by basing this
on binary trees, and later proceed to the multiway tree.  An alternative
approach can be found in Ganesan et al.\ [2003], where similarities or 
distances on trees are redefined and re-axiomatized for the case of 
multiway trees.  
%P. Ganesan, H. Garcia-Molina and J. Widom, ``Exploiting hierarchical
%domain structure to compute similarity'', ACM Trans on IS, 21, 64-93,
%2003.
%Defines sim or distance on trees, where trees are multiway, or with 1
%offspring, and otherwise not binary.

An alternative representation for an ontology is a lattice, and Formal 
Concept Analysis (FCA) is a methodology for the analysis of such lattices.  
If we have a set of documents or texts, $I$, characterized by an 
index term set $J$, then as Janowitz [2005] shows, hierarchical clustering
 and FCA are loosely related.  
Hierarchical clustering is based on pairwise distances
or dissimilarities, $d: I \times I \rightarrow \R^+$ ($\R^+$ is the set of 
non-negative reals).  FCA is based on partially ordered sets (posets) such 
that there is a dissimilarity  $d: I \times I \rightarrow 2^J$ ($2^J$ is the 
power set of the index terms, $J$). 

% Should I go into: concept hierarchy = taxonomy   = ? thesaurus?

Other approaches (rule-based; machine learning approaches, etc.; layered, 
engineering, approaches with maintenance management -- see Maedche [2006])
are also available.  One difficulty with such ``engineering'' 
approaches is that
there is an ad hoc understanding of the problem area, and often there is 
dependence on somewhat arbitrary threshold and selection criteria that do 
not generalize well.  

Our approach formalizes the problem area -- the 
information space -- in terms of its local or global topology.  Where 
we do have selection criteria, such user interaction
is  at the application goal level.  

Visualization is often an important way to elucidate semantic heterogeneity 
for the user.  Visual user interfaces for ontological elucidation are 
discussed in Murtagh et al.\ [2003], with examples that include 
interactive, responsive information maps based on the Kohonen self-organizing
feature map; and semantic network graphs.  A study is presented in Murtagh 
et al.\ [2003] of client-side visualization of concept hierarchies relating
to an  economics information space.  The use of ``semantic road maps'' 
to support information retrieval goes back to Doyle [1961].   Motivation,
following Murtagh et al.\ [2003], includes the following: 
(i) Visualization of the semantic
structure and content of the data store allows the user to have some idea
before submitting a query as to what type of outcome is possible.  Hence
visualization is used to summarize the contents of the database or data
collection (i.e., information space).  (ii) The user's information 
requirements are often fuzzily and ambiguously 
defined at the outset of the information 
search.  Hence visualization is used to help the user in his/her information
navigation, by signaling the relationships between concepts.  (iii) 
Ontology visualization therefore helps the user before the user interacts 
with the information space, and during this interaction.  It is a natural 
enough progression that the visualization becomes the user interface.

%In the work presented here, we use a part of speech tagger to locate nouns
%for further analysis.  However {\em all} words in the documents used are
%retained in the analysis, since the aggregate of all pairwise relationships
% serves to locate any given term.  Terms are demarcated by white space or 
%punctuation.  Stemming is not used, thereby allowing both singulars and 
%plurals (for example) to carry potentially different semantics (cf.\ Murtagh,
%2005b).    Given that frequency of occurrence, including presence/absence,
%information is a standard starting point for text analysis, we use the 
%$\chi^2$ metric, incorporating row and column weighting, and map this into 
%a Euclidean output space, called a factor (or principal component) space.
%This is done using correspondence analysis and justifies use of the Euclidean 
%embedding at a key point in the analysis pipeline.  

\subsection{Organisation of the Article}

This article is organised as follows.  To begin with,
in section \ref{sect2},  we table the issue of 
whether or not there is inherent hierarchical structure in a text, or a
collection of texts.  In section \ref{sect3} 
 we show how we can rigorously determine the extent of 
inherent hierarchical structure in a text.  
This quantifying of inherent hierarchical structure is then used in 
subsections \ref{sect41}, \ref{sect42}, \ref{sect52} and \ref{sect7}. 

A text provides both global and linear semantics, and how we can process 
these two different perspectives on a given text is discussed in 
section \ref{sect4}.
A central aspect of our approach is a new distance or metric, which we
have recently introduced and exemplified on another data analysis problem.
This new distance is described in subsection \ref{sect5}.

In section \ref{sect6} we apply what we have described in earlier sections
to the selection of salient and characteristic pairs and triplets of terms,
and also the selection of pertinent terms.  Our motivation is not just 
the traditional view of phrase counting (even though we incorporate this 
view)  but rather the characterization
of text content using its internal (local hierarchical) structure.  

A natural approach to defining a concept hierarchy lies in use of a 
hierarchical clustering algorithm.  However, the latter forms an embedded
sequence of clusters, so that a hierarchy of concepts must -- somehow -- 
be derived from it.  
In section \ref{secorientree} we first of all show that ``converting'' any 
hierarchical clustering into a hierarchy of concepts is relatively 
straightforward.  However we do have to face the problem of a 
unique, and beyond that best, solution.  We show how we can admirably 
address this need for a unique solution.  Our innovative approach is 
based on the foundations laid in sections \ref{sect2} and \ref{sect4} 
of this article.

We analyze three different data sets in this work: firstly a set of 
documents, with some degree of heterogeneity,
to illustrate our key goal; secondly a homogeneous 
text, partitioned into successive textual segments; and thirdly a small
homogeneous text, partitioned at the sentence level, proxied by lines of
text.  We select terms, indeed nouns, in a partially automated way, 
since this crucial aspect of ontology design may benefit from being
 user-driven, and may have scalability advantages.

\section{Quantifying Hierarchical Structure}
\label{sect2}

In later sections we address the issue of finding and presenting 
structure in text.  We link such structure with the textual content.  
Consequently a key, initial question is to know whether or not there
is structure present, and to what extent.  

\subsection{Inherent Hierarchical Structure}

A first problem to be addressed is  whether or not the 
document has any hierarchical structure to begin with.
As input, we have possibly a fully tagged document (based, e.g., on 
part-of-speech tagging, Schmid [1994]).  
However in this work, we start with free text, because it is the 
most generally available and applicable framework.  
%Our point of departure
%is the set of words or stemmed terms comprising the document.  
Additional information provided by part-of-speech tagging 
can be of use to us,  as we will show later.

Next we consider the issue of whether or not a document has 
sufficient inherent hierarchical structure to warrant further 
investigation.  We could approach this problem by fitting a hierarchy,
and there are many algorithms for doing so (such as any hierarchical clustering
algorithm; de Soete [1986] describes a least squares optimal fitting 
approach).  However departure from inherent hierarchical structure is 
not easily pinpointed.  After all, we have an output induced structure, 
and we are told, let's say, that the fit is 80\%  (defined as 
$\sum (\delta - d)^2/\sum d^2$ where $d$ is input dissimilarity, 
$\delta$ is tree or ultrametric distance read off the output, and the 
sums are over all pairs), which is not very revealing nor useful.  

An alternative ``bottom-up'' approach is pursued here, which allows
easy assessment of inherent structure, and also pinpointing where 
this occurs or does not occur.  

\subsection{Local Ultrametricity and Quantifying Extent of Ultrametricity}
\label{sect3}

A formal definition of hierarchical structure is provided by ultrametric
topology (in turn, related closely  to p-adic number theory). 
The triangular inequality holds for a metric space: $d(x,z) \leq
d(x,y) + d(y,z)$ for any triplet
of points $x,y,z$.  In addition the properties
of symmetry and positive definiteness are respected.  The ``strong
triangular inequality'' or ultrametric inequality is: $d(x,z) \leq
\mbox{ max } \{ d(x,y), d(y,z) \}$ for any triplet $x,y,z$.  An
ultrametric space implies respect for a range of stringent properties. 
 For example, the triangle formed by any triplet is necessarily isosceles, 
with the two large sides equal; or is equilateral.  In an ultrametric
space (i.e., a space endowed with an ultrametric, or an ultrametric 
topology), one ``lives'', so to speak, in a tree.  All ``moves'' between 
one location and another are as if one descended the tree to a common 
tree node, and then reclimbed to the target point.  Topologically,
an ultrametric goes a lot further: all points in a circle or sphere are
centers, for example; or the radius of a sphere is identical to its
diameter.  

The triangle property respected by any triplet of points in 
an ultrametric space affords a useful way to quantify extent of 
hierarchical structure.  We will describe our ``extent of hierarchical
structure'', on a scale of 0 (no respect for ultrametricity) to 1
(everywhere, respect for the ultrametric or tree distance)  
algorithmically.  We 
examine triplets of points (exhaustively if possible, or otherwise
through sampling), and determine the three angles formed by the
associated triangle.  We select the smallest angle formed by the triplet
points.  Then we check if the other two remaining angles are
approximately equal.  If they are equal then our triangle is isosceles
with small base, or equilateral (when all triangles are equal).  The
approximation to equality is given by 2 degrees (0.0349 radians).
Our motivation for the approximate (``fuzzy'') equality is that it 
makes our approach robust and independent of measurement precision.
This approach works very well in practice [Murtagh 2004; 2006a].  We 
may note our one assumption for our data when we look at triangles in this
way: scalar products define angles so that by assuming our data are in 
a Hilbert space (a complete normed vector space with a scalar product) we
may proceed with this analysis.  This Hilbert space assumption is very 
straightforward in practice.  When finite (as is always the case for 
us, in practice), we are using a Euclidean space.  

Often in practice, for arbitrary Euclidean data, there is very little 
ultrametricity as quantified by the proportion of triangles satisfying the 
ultrametric requirement.  But recoding the data can be of great help in 
dramatically increasing the proportion of such ultrametricity-respecting
triangles [Murtagh 2004; 2005a].  
If we recode our data such that each pairwise distance or dissimilarity is 
mapped onto one element of the set $\{0, 1, 2\}$, then as seen in subsection 
\ref{sect5}
below the triangular inequality becomes particularly easy to assess 
for existence of, or non-existence of, a locally ultrametric relationship.

\section{Global and Linear Structures of a Text}
\label{sect4}

\subsection{Euclidean Embedding}

In our use of free text, we have already noted how a mapping into a 
Euclidean space gives us the capability to define distance in a simple
and versatile way.  In correspondence analysis [Murtagh 2005b], the texts 
we are using provide the rows, and the set of terms used comprise the 
column set.  In the output, Euclidean factor coordinate space, each text is 
located as a weighted average of the set of terms; and each term is located
as a weighted average of the set of texts.  (This simultaneous display is 
sometimes termed a biplot.)  So texts and terms are both 
mapped into the same, output coordinate space.  This can be of use in 
understanding a text through its closest terms, or vice versa.  

A commonly used methodology for studying a set of texts, or a set of 
parts of a text (which is what we will describe below), is to characterize each
text with numbers of terms appearing in the text, for a set of terms.
The $\chi^2$ distance
is an appropriate weighted Euclidean distance for use with such data
[Benz\'ecri 1979; Murtagh 2005b].
Consider texts $i$ and $i'$ crossed by words $j$.  Let $k_{ij}$ be the
number of
occurrences of word $j$ in text $i$.  Then, omitting a constant,
the $\chi^2$ distance between texts $i$ and $i'$ is given by
$ \sum_j 1/k_j ( k_{ij}/k_i - k_{i'j}/k_{i'} )^2$.  The weighting term is
$1/k_j$.  The weighted Euclidean distance is between the {\em profile}
of text $i$, viz.\ $k_{ij}/k_i$ for all $j$, and the analogous
{\em profile} of text $i'$.  (Our discussion is to within a constant because
we actually work on {\em frequencies} defined from the numbers of occurrences.)

Correspondence analysis allows us to project the space of documents (we could
equally well explore the terms in the {\em same} projected space) into
a Euclidean space.  It maps the all-pairs $\chi^2$ distance into the
corresponding Euclidean distance. 

For a term, we use the (full rank) projections on factors resulting from
correspondence analysis.  As noted, this factor space is endowed with the
(unweighted) Euclidean distance.  

\subsection{Linearity: Textual Time Series}

We will also take into consideration the 
strongest ``given'' in regard to any classical text: its linearity (or 
total) order. A text is read from start to finish, and consequently is 
linearly ordered.  
%(Informally we will refer to the text as being ``linear'').  

A text endowed with this linear order is analogous to a time series.  
%(This opens up the possibility to generalize the work described here 
%to (i) speech signals, or (ii) music.  We will pursue these
%generalizations in the future.) 
If we use the correspondence analysis (full dimensionality) factor 
coordinates for each term, then the textual time series we are dealing 
with is seen to be a multivariate time series.  

\subsection{Recoding Distances}
\label{sect5}

Just as the way we code our input data plays a crucial role in the
resulting analysis, so also the recoding of pairwise distances can influence
the analysis greatly.  In Murtagh [2005a] we introduced a new distance,
which we will term the ``change versus no change'', CvNC, 
metric, and showed its
benefits on a wide range of (financial, biomedical, meteorological, telecoms,
chaotic, and random) time series.  Motivation for using this new metric
is that it greatly increases the ultrametricity of the data.  

The CvNC metric is defined in the following way.  Take the 
Euclidean distance squared, $d_{jj'} = (x_{jr} - x_{j'r})^2$
for all $ 1 \leq j, j' \leq m$, where we have terms $j, j'$ in the factor 
space with coordinates $1, 2, \dots , r, \dots \nu$.  
It will be noted below in this section
how this assumption of Euclidean distance squared has worked well but is
not in itself important: in principle any dissimilarity can be used.

We enforce sparseness on our given squared distance
values, $\{ d_{jj'} \}$.  We do this by approximating
each value $d_{jj'}$, in the range $ \mbox{max}_{jj'} d_{jj'} -
\mbox{min}_{jj'} d_{jj'} $, by an integer in $1, 2, \dots p$.  The value of
$p$ must be specified.  In our work we set  $p = 2$.  The recoding of
distance squared is with reference to the mean distance squared: values 
less than or equal to this will be mapped to 1; and values greater than 
this threshold will be mapped to 2.  Thus far, the recoded value, $d'_{jj'}$
is not necessarily a distance.  With the extra requirement that
$d'_{jj'} \longrightarrow 0$ whenever $j = j'$ it can be shown that
$d'_{jj'}$ is a metric [Murtagh 2005a]:

\bigskip

{\bf Theorem:} {\em The recoded pairwise measure, $d'$, defined as described 
above from any dissimilarity, is a distance, satisfying the properties
of: symmetry, positive definiteness, and triangular inequality.}

\bigskip

To summarize, in our coding,
a small pairwise dissimilarity is mapped onto a value of
1; and a large pairwise dissimilarity is mapped onto a value of
2.  Identical values are unchanged: they are mapped onto 0.

This coding can be considered as 
encoding pairwise relationships as ``change'', i.e.\ 2, versus
``no change'', i.e.\ 1, relationships.  Then, based on these new distances,
we use the ultrametric triangle properties to assess conformity to
ultrametricity.   The proportion of ultrametric triangles 
allows us to fingerprint our data.

For any given triplet (of terms, with pairwise CvNC distances), 
if the triplet is to be compatible with 
the ultrametric inequality, each set of three recoded distances is necessarily
of one of the following patterns: 

\begin{description}
\item[Trivial:] 
At least one (recoded) distance is 0, in which case we do not consider
it.
\item[Ultrametric -- equilateral:]  Recoded distances in the triplet 
are  1,1,1 or 2,2,2, defining an equilateral triangle.
\item[Ultrametric -- isosceles:]
Recoded distances in the triplet are 
1,2,2 in any order, defining  an isosceles triangle with small base. 
\item[Non-ultrametric:] Recoded distances in the triplet are
1,1,2 in any order. 
\end{description}

The non-ultrametric case here is seen to be an isosceles triangle with 
large base.  We could ``intervene'' and change one of the values to make 
it ultrametric.  If we change the 2-value to a 1-value, this will produce
an equilateral triangle, which is ultrametric.  In this case, we are 
approximating our three values optimally from below, and the resulting 
ultrametric is termed the subdominant, or maximally inferior,
 ultrametric.  The associated 
stepwise algorithm for constructing a hierarchy is known as the single 
link hierarchical clustering algorithm.  On the other hand, we could change one
of the 1-values to a 2.  This is not unique, since we could change either 
of the 1-values.  The resulting hierarchy is termed the minimally superior 
ultrametric.  The associated
stepwise algorithm for constructing a hierarchy is known as the complete
link hierarchical clustering algorithm.  All of this is very clear from the 
case considered here.  

The recoding into the CvNC metric is a particular example of symbolic 
coding.  See Murtagh [2006c].  

In the next section, we will show the usefulness of this CvNC metric 
for quantifying inherent hierarchical structure.

\section{Application to Pair and Triplet Phrase 
Finding, and to Selecting Pertinent Terms}
\label{sect6}

In this section we first describe the data set used.  
Next, based on the foundation
of the previous section, we quantify inherent hierarchical structure in 
our data.  
This justifies going further, to harness and exploit this structure.  

\subsection{Data Used}
\label{sect41}

We use 14 texts taken from Wikipedia (mid-2006), 
and coverted to straight text from 
HTML.  Table \ref{tabprop} shows the numbers of words in each. 

\renewcommand{\baselinestretch}{1}

\begin{table}
\begin{center}
\begin{tabular}{llrrr}\hline 
File    &    Theme                &        No. &    No. &  No.  \\
        &                         &        words &  nouns &  uniq. nouns \\
 \hline 
        &                         &             &         &    \\
ArtInt  &   Artificial Intelligence   &    1624 &    405  &  231 \\
ArtLife &   Artificial Life           &    2095 &    448  &  275 \\
ArtNN   &   Artificial Neural Network &    4698 &   1262  &  389 \\
Captcha &   Captcha                   &    1479 &    318  &  169 \\
CompLin &   Computational Linguistics &     648 &    168  &   80 \\
CompVis &   Computer Vision           &    2396 &    737  &  311 \\
EvolCom &   Evolutionary Computation  &     156 &     58   &  43 \\
FuzzLog &   Fuzzy Logic               &    1663 &    399  &  204 \\
GenAlg  &   Genetic Algorithms        &    2775 &    715  &  306 \\
MaTrans &   Machine Translation       &    1643 &    411  &  172 \\
MAgent  &   Multi-agent System        &     493 &    104  &   67 \\
SemNet  &   Semantic Network          &     475 &     96  &   74 \\
Turing  &   Turing Test               &    2432 &    459  &  225 \\
VirtW   &   Virtual World             &     583 &    144  &   79 \\
        &                             &         &         &      \\
All files &                           &         &   5724 &  1165 \\ \hline  
\end{tabular}
\end{center}
\caption{Properties of texts used.}
\label{tabprop}
\end{table}

\renewcommand{\baselinestretch}{2}

%MaLearn    Machine Learning              2775     715    306
%All files                                        6439   1471     

% sort TTout15_nouns.txt  | wc 
% sort TTout15_nouns.txt  | uniq | wc
% sort TTout*_nouns.txt | wc 
% sort TTout*_nouns.txt | uniq | wc 

We derived 4048 unique terms (all parts of speech, including nouns)
from the collection of 14 texts listed in Table \ref{tabprop}.  
As noted before, we do not apply stemming.  The $14 \times 
4048$ frequency of occurrence matrix was analyzed using correspondence 
analysis, which furnished an embedding of both texts and terms in a 
13-dimensional (i.e., necessarily at most one less than the minimum of 
input row and column dimensions, viz.\ 14 and 4048) factor space.  

Although the presence in the analysis of minor words can be important
(see discussion in Murtagh [2005b]),
for the concept hierarchy relationships we are primarily interested in 
nouns.  We used therefore a part of speech tagger [Schmid 1994]
to locate the nouns.  The number of nouns found in the Artificial 
Intelligence text  was 231.  For each we have
a 13-dimensional factor space representation, and the latter has been
 defined globally, using all texts in the collection studied.  

For all pairs of the these 231 terms, using their 13-dimensional 
Euclidean characterization, we carried out the mapping into the CvNC 
metric.  

For the Artificial Intelligence text, 36\% of the triplets were 
equilateral; 50\% of the  triplets were isosceles with small base; 
hence 86\% of the triplets respected the ultrametric inequality.  
Finally, 14\% of the triplets were non-ultrametric.  

A summary of the ``fingerprinting'' procedure in regard to the text's 
ultrametricity, or inherent local hierarchical structure, is as follows.

\begin{enumerate}
\item Define each of the relevant terms -- nouns -- in a Euclidean 
factor space.  
\item Take each triplet of terms in turn.
\item Define the squared Euclidean distance between each
successive pair of terms.   
\item Use the pairwise average of these squared distances as a threshold.
\item If the pair of terms is of squared distance less than the threshold,
then define their relationship as ``no change''.
\item If the pair of terms is of squared distance greater than or equal
to the threshold, then define their relationship as ``change (either up
or down)''.
\item With ``no change'' coded as 1, ``change'' coded as 2, and
self-distances coded as 0, Murtagh [2005a] shows that the resulting
mapping of the Cartesian product of terms $\times$ terms onto the set
$ d' \in  \{ 0, 1, 2 \}$ defines a metric.  For all terms $i, j, k$, we
therefore have $d'_{ij} \leq d'_{ik} + d'_{kj}$.
\item For the given triplet we check if this metric is an ultrametric:
 For terms $i, j, k$, we
therefore seek whether $d'_{ij} \leq \mbox{max} \{
d'_{ik}, d'_{kj} \}$.
\item If the triplet $i, j, k$ respects the ultrametric relation, then
there are two possible cases.  Firstly, the triangle formed by these terms
is equilateral, which is implied whenever $d'_{ij} = d'_{ik}
= d'_{kj}$.  Secondly, the triangle is isosceles with small base,
which is implied by two $d'$s being equal, and greater in value to
the third.
\item No other triangle configurations are consistent with the ultrametric
relationship.
\item Over all triplets considered, the ultrametricity index of the 
document is the proportion of ultrametricity-respectiving triplets.  
\end{enumerate}

Table \ref{tabUMwiki} shows the results obtained for (i) global relationships 
given by all triangles (triangles
were read off using the loops  
$i=1 \dots n-2$,  $j=i+1 \dots n-1$, $ k = j+1 \dots n$); 
and (ii) linear relationships, using only triangles defined from 
successives triplets
of terms in the text.  Nouns were used: cf.\ Table \ref{tabprop}.  
What we see very
clearly  from Table \ref{tabUMwiki} is that, whether global or linear, 
our texts 
show very dominant ultrametric or hierarchical structure.  Furthermore, 
this is, in 
the great majority of cases, dominated by the isosceles with small base 
case, relative
to the ``trivial'' equilateral case.  

These results justify going further now, in order to make use of the inherent 
hierarchical structure that is in our data.  

\renewcommand{\baselinestretch}{1}

\begin{table}
\begin{center}
\begin{tabular}{lrrrr} \hline
Text             & Total no. & No. isosc. & No. equil. & Non-UM \\
                 & triangles & triangles  & triangles  & triangles \\ \hline
                 &           &            &            &        \\
Artificial Intelligence   & 2027795 & 50 & 36 & 14 \\
                          & 336     & 37 &  45 & 18 \\
Artificial Life           & 3428425 & 52 &  36 & 12 \\
                          & 342     & 34 &  41 & 26 \\
Artificial Neural Network & 9735114 & 46 &  42 & 12 \\
                          & 914     & 39 &  49 & 11 \\
Captcha                   & 790244  & 54 &  33 & 12 \\
                          &  225    & 41 &  33 & 25 \\
Computational Linguistics & 82160   & 46 &  39 & 15 \\
                          & 132     & 47 &  41 & 12 \\
Computer Vision           & 4965115 & 50 &  38 & 13 \\
                          &  578    & 36 & 40 &  24 \\
Evolutionary Computation  & 12341   & 37 & 48 & 15 \\
                          &  52     & 33 & 54 & 13 \\
Fuzzy Logic               & 1394204 & 52 & 33 & 14 \\
                          &  299    & 45 & 30 &  25 \\
Genetic Algorithms        &  4728720 & 43 & 46 & 10 \\
                          &  581    & 37 & 51 & 12 \\
Machine Translation       & 833340  & 49 & 36 & 15 \\
                          & 331     & 40 & 33 &  27 \\
Multi-agent System        & 47905   & 49 & 42 & 9 \\
                          & 87      & 37 & 49 & 14 \\
Semantic Network          & 64824   & 59 &  35 & 6 \\
                          & 77      & 35 & 57 &  8 \\
Turing Test               & 1873200 & 50 & 37 & 13 \\
                          & 365     & 45 & 33 & 23 \\
Virtual World             & 79079   & 58 & 29 &  13 \\
                          & 88      & 42 &  28  & 30 \\ \hline
\end{tabular}
\end{center}
\caption{Texts, and their properties as quantified, as percentages,
 by relative numbers of
triangles.  Note the percentages in columns 3, 4 and 5 sum to 100, but 
may deviate from this by a unit due to the rounding to the nearest integer.
For each text, we show results for  the global and the linear cases (see
accompanying discussion for details).}
\label{tabUMwiki}
\end{table}

\renewcommand{\baselinestretch}{2}

\subsection{Application to Concept Hierarchy Relationships}
\label{sect42}

To the extent that our data satisfies, globally and throughout,
 the ultrametric inequality, we can adopt any of the widely used 
hierarchical clustering algorithms (single, complete, average linkage;
minimum variance, median, centroid) to induce an identical, unique
hierarchy.  But when we find our data to be, say, 86\% ultrametric, as is 
not untypically the case in practice, then we must consider carefully what 
our aim is.  If we wished to look at each and every isosceles triangle,
then in the case of the Artificial Intelligence text this means, out of
a total of 2,027,795 triplets (i.e., $(231 \cdot 230 \cdot 229) / (2 \cdot 3)$)
we must consider 1,007,597.   

What we will do instead is return to taking our text as a time series. 
We have 231 unique nouns in the Artificial Intelligence text.  
In the text, these nouns are used, in total, 
on 405 occasions.  So our text is a time series of 405 values.  For 
successive nouns in this textual time series, the CvNC metric has an evident 
 meaning: {\em we are noting semantic change versus lack of change 
as we read through the text}.  

We examine successive triplets in the textual time series.  For the 
Artificial Intelligence text, we find 45\% of the triplets to be 
equilateral; 37\% of the triplets are isosceles; and 18\% of the triplets
are non-ultrametric.  

The isosceles triplets point to a dominance or subsumption relationship
that will be of use for us in a concept hierarchy.  Say we have a triplet
$x, y, z$.  Say, further, that the CvNC distance between $x$ and $y$ is 1,
so therefore there is no change in progressing from use of term $x$ to 
use of term $y$.  However both $x$ and $y$ are at CvNC distance 2 to term
$z$, and this betokens a semantic change.  
So the relationship is simply represented as $((x, y) z)$.  The term $z$
dominates or subsumes $x$ and $y$.

The following results hold.  

Firstly, say that a successive triplet of values,
in any order, is found as $x, y, z$, and later in the text, again, this 
triplet is found in any order.  Then the relationship between the three
recoded distances in both cases will be identical.  {\em For a given triplet,
in any order within the triplet, the relationship is unique.}  

Secondly, consider any other term, $w$, such that some or all of the 
terms $x, y, z$ are found to have a relationship with $w$.  As an example,
we meet with $y, z, x$ at one point in the text, and later we meet with
$x, w, y$.  Then there is no influence by $w$ on  the relationship 
ensuing from the $x, w, y$ triplet, vis-\`a-vis the relationship ensuing 
from the earlier $y, z, x$ triplet.  {\em We have locality of 
the relationship in any given triplet, from successive terms.
The relationship is strictly local to the given triplet.}

Among the isosceles triangles in the Artificial Intelligence text, 
we find the following relationships.  

\begin{verbatim}
( computer science ) branch 
( home computer ) world 
( analysis systems ) formalism 
( analysis systems ) reasoning 
( expert system ) conclusion 
( expert system ) amounts 
( example networks ) reasoning 
( networks learning ) reasoning 
( pattern recognition ) capabilities 
( control systems ) computation 
( consciousness systems ) logic 
( medicine computer ) commentators 
( computer technology ) commentators 
( application feature ) os 
( application feature ) languages 
( libraries systems ) specialist 
( libraries systems ) programmers 
( software engineering ) development 
( software engineering ) practices 
( programs example ) logic 
( example type ) logic 
( projects publications ) life 
( publications bayesian ) life 
( bayesian networks ) life 
( cybernetics systems ) agents 
( systems control ) agents 
( wiki web ) website 
( wiki web ) category 
( algorithm implementations ) projects 
( algorithm implementations ) demonstrations 
( implementations research ) demonstrations 
( research group ) demonstrations 
\end{verbatim}

However there are other isosceles triplets that are less self-evident.  
For this reason therefore we  take all texts.  For the 14 texts,
we have  %using AllxTToutxnouns.txt
6439 nouns, and 1470 unique nouns.  
With our CvNC metric on all pairs of nouns, the complete link hierarchical
clustering method gives 21 clusters in all.  

%Single link of 1/2 data: 1 cluster with 2 elements, classifier, swarms, 
%and all remaining elements were in other cluster.  
%Complete link better: 
%By extending the screen, we see that ``prototype'', ``matter'' are the
%# only 2 terms merged in at level 2 for res2;
%# And we have non-empty clusters up to cluster 21.
%res2p[res2p==21]
%classifier     swarms
%        21         21
%Now select set of 73 terms.  

While one application of the foregoing is to deriving common pairs and 
triplets of terms, in practice it would be better to combine all 
relationships into a ``bigger picture''.  We will address this below in 
section \ref{secorientree}. 

\subsection{Selecting the Most Pertinent Terms}

\begin{figure}
\includegraphics[width=14cm]{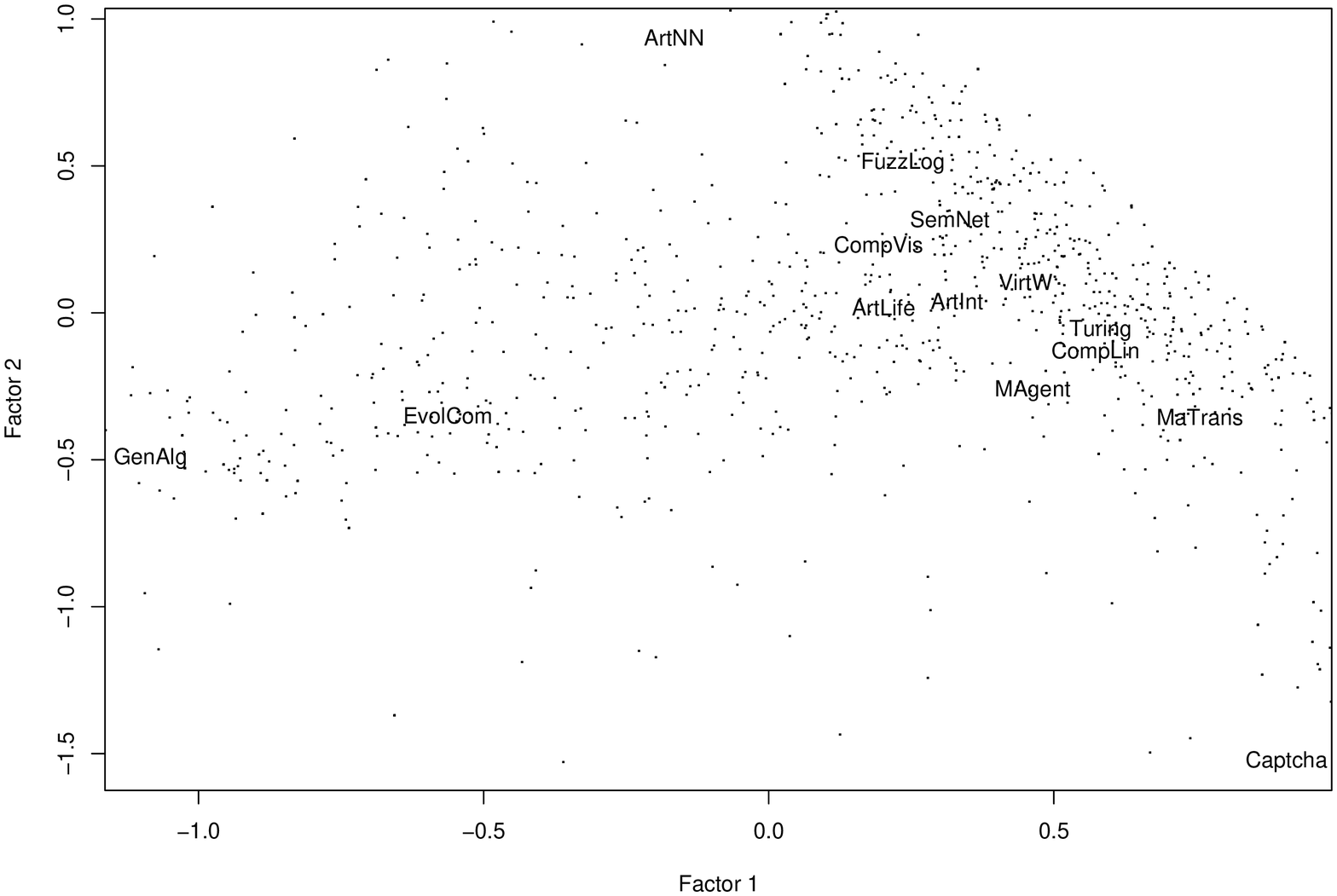}
\caption{Correspondence analysis principal factor plane of projections 
of 14 texts, and 4048 terms (each represented with a dot).}
\label{caout}
\end{figure}

\begin{figure}
\includegraphics[width=10cm,angle=270]{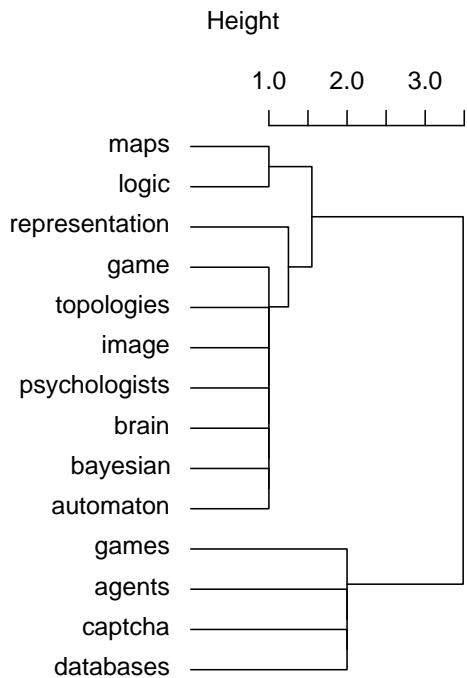}
\caption{Dendrogram on the 14 noun-set, selected as the set of  closest 
terms to each of the 14 texts used.  
The hierarchical clustering is based on a full dimensionality
correspondence analysis factor embedding.  Ward's minimum variance 
clustering is used.}
\label{den15}
\end{figure}

Presenting a result with around 1500 terms does not lend itself to 
convenient display.  We ask therefore what the most useful -- perhaps the 
most discriminating terms -- are.  In correspondence analysis both texts 
and their characterizing terms are projected into the same factor space.
See Figure \ref{caout}.   
So, from the factor coordinates, we can easily find the closest term(s) 
to a given text.  We do this for each of the 14 texts, and find the closest
terms, respectively, as follows:

\begin{verbatim}
     bayesian         automaton    brain          captcha
     psychologists    image        maps           logic
     topologies       databases    agents         representation   
     game             games
\end{verbatim}

A hierarchical clustering of these is shown in Figure \ref{den15}.  
The Ward minimum variance method is used, as being appropriate for
structuring data well (see Murtagh [1984b]) and also having an agglomerative
criterion that is appropriate for the prior Euclidean embedding (viz., 
inertia-based in both cases).  The data clustered are exactly those
illustrated in the best planar projection of Figure \ref{caout}: these
are 14 texts in a 4048-dimensional term space.  Due to centering in 
the dual spaces, the inherent dimensionality of both text and term 
spaces are: min$(14 - 1, 4048 - 1)$ = 13.  Based on the dual spaces, 
we carry out the eigen-reduction  in the space of smaller
original dimensionality (viz., the space of the terms, which are in a 
14-dimensional space), and then subsequently project into the 
4048-dimensional space.  

Proceeding further, the 5 closest terms to any given text, based on the 
full inherent dimensionality of this data (viz., smaller of dimensionality 
of texts, and dimensionality of terms), are as follows.  

\begin{verbatim}
Text and set of 5 closest characterizing terms:

Artificial Intelligence 
bayesian intelligence consciousness brains chatterbots

Artificial Life
automaton automata biology chemical allelomimesis

Artificial Neural Networks
brain prediction forecasting aircraft epitomes

Captcha
captcha captchas robot intelligence chemistry

Computational Linguistics
psychologists logics morphology pragmatics logicians

Computer Vision
image images diagnosis dimensionality dimensions

Evolutionary Computation
maps intelligence robot biology chemistry

Fuzzy Logic
logic mapping animals brakes armies

Genetic Algorithms
topologies communications music finance representations

Machine Translation
databases chemistry database memory robot

Multi-agent System
agents agent robotics cybernetics robot

Semantic Network
representation database map namespaces robot

Turing Test
game chatterbot consciousness memory intelligence 

Virtual World
games gameplay topography communication representations
\end{verbatim}

\section{From Hierarchical Clustering to a Hierarchy of Concepts}
\label{secorientree}

\subsection{A Formal Approach: Displaying a Hierarchical Clustering
as an Oriented Tree}

% See wiki-work/wiki_ai audit-orientedtree.txt

We have noted in the Introduction how a hierarchical clustering may be 
the starting point for creating a concept hierarchy, but the two 
representations differ.  In this section we show how we can move from an
embedded set of clusters, to an oriented tree.  Orientation in the latter
case aims at expressing subsumption.

\begin{figure*}
\begin{center}
\includegraphics[width=14cm]{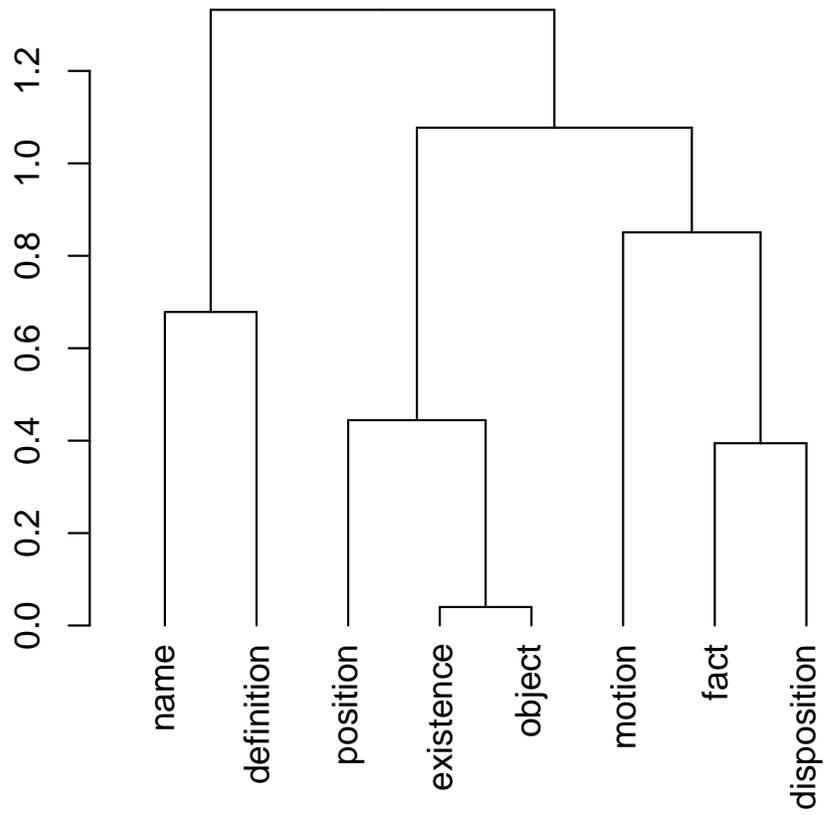}
\end{center}
\caption{Hierarchical clustering of 8 terms.  Data on which this
was based: frequencies of occurrence of 66 nouns in 24 successive,
non-overlapping segments of Aristotle's {\em Categories}.}
\label{den1}
\end{figure*}

Consider the dendrogram shown in Figure \ref{den1}, which represents an
embedded set of clusters relating to the 8 terms.  We will consider first 
such a strictly 2-way hierarchy, where we assume that no two agglomerations
take place at precisely the same level.  In the later case study, in 
subsection 
\ref{sect7}, we will consider the practical case where agglomerations take 
place at the same level.  
  
Rather than the 14 texts used in section \ref{sect6}, to clarify 
the presentation in this section we will take just one text. 

 We took Aristotle's
{\em Categories}, which consisted of 14,483 individual words.  We broke
the text into
24 files, in order to base the textual analysis on the sequential properties
of the argument developed.  In these 24 files there were 1269 unique words.
We selected 66 nouns of particular interest.  A sample (with
frequencies of occurrence):
man (104), contrary (72), same (71), subject (60), substance (58), ...
No stemming or other preprocessing was applied.  For the hierarchical
clustering, we further restricted the set of nouns to just 8.  (These will
be seen in the figures to be discussed below.)
The data array was
doubled [Murtagh 2005b] to produce an $8 \times 48$ array, which with
removing 0-valued text segments (since, in one text segment,
 none of our selected 8 nouns appeared) gave an $8 \times 46$ array,
thereby
enforcing equal weighting of (equal masses for) the nouns used.
The spaces of the 8 nouns, and of the 23 text segments (together with the
complements of the 23 text segments, on account of the data doubling)
are characterized at the start of the correspondence analysis in terms of their
frequencies of occurrence, on which the $\chi^2$ metric is used.  The
correspondence analysis then ``euclideanizes'' both nouns and text segments.
%Such a Euclidean embedding is far safer for later processing, including
%clustering (and frankly would be most ad hoc, and/or ``customized'' and
%less general, in terms of any alternative data analysis).  
We used a
7-dimensional (corresponding to the number of non-zero eigenvalues found)
Euclidean embedding, furnished by the projections onto the
factors.  A hierarchical clustering of the 8 nouns, characterized by their
7-dimensional (Euclidean) factor projections, was carried out: Figure
\ref{den1}.  The Ward minimum variance agglomerative criterion
was used, with equal weighting of the 8 nouns.
 
\begin{figure}
\includegraphics[width=14cm]{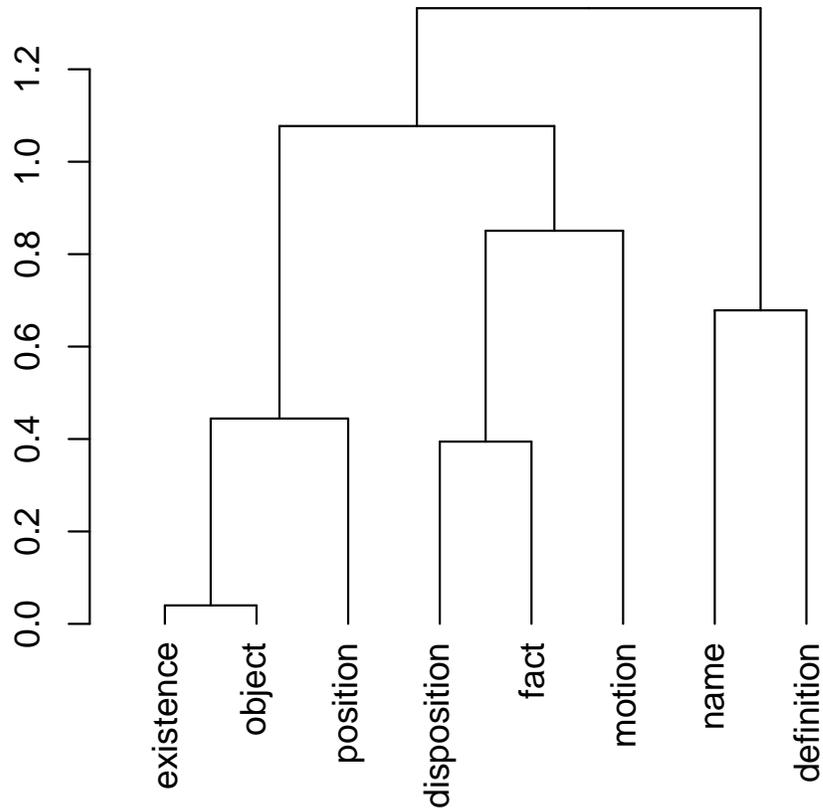}
\caption{Dendrogram on 8 terms, isomorphic to the 
 previous figure, Figure \ref{den1},
but now with successively {\em later}
agglomerations always represented by {\em right} child node.  Apart from the 
labels of the initial pairwise agglomerations, this is otherwise a unique 
representation of the dendrogram (hence: ``existence'' and ``object''
can be interchanged; so can ``disposition'' and ``fact''; and finally 
``name'' and ``disposition'').  In the discussion we refer to 
this representation, with later agglomerations always parked to the right,
as our canonical representation of the dendrogram.}
\label{den2}
\end{figure}

\begin{figure}
\includegraphics[width=14cm]{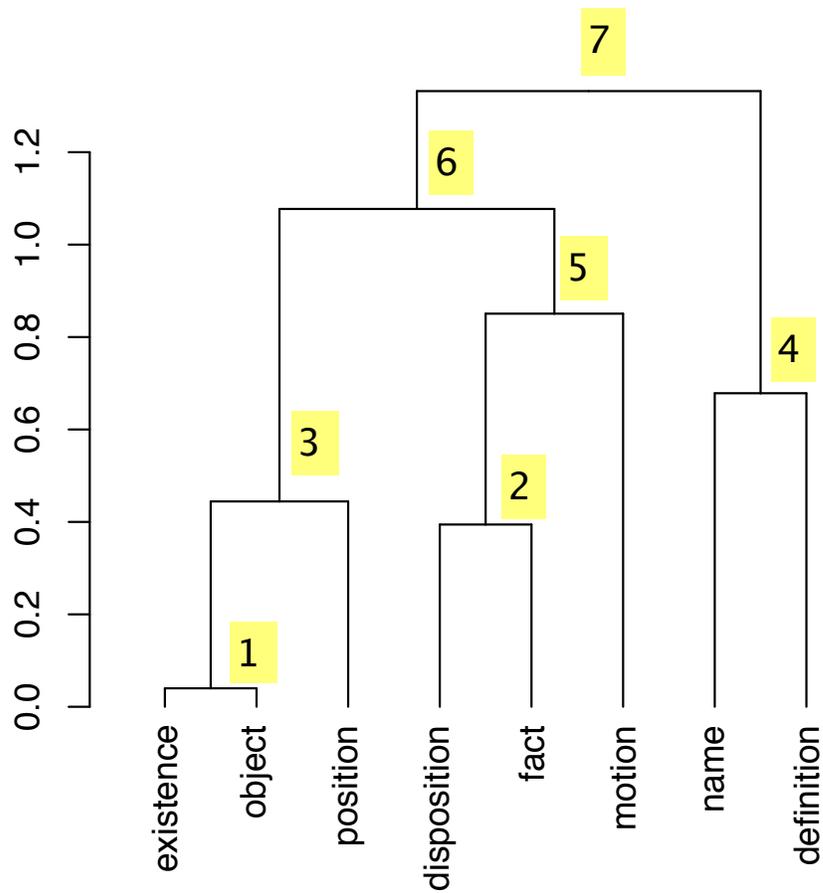}
\caption{Dendrogram on 8 terms, as previous figure, Figure \ref{den3},
with non-terminal
nodes numbered in sequence.  These will form the nodes of the oriented
binary tree.  We may consider one further node for completeness, 
8 or $\infty$, located 
at an arbitrary location in the upper right.}
\label{den3}
\end{figure}

Figure \ref{den2} shows a canonical representation of the dendrogram in 
Figure \ref{den1}.  Both trees are isomorphic to one another.
Figure \ref{den2} is shown such that the sequence of 
agglomerations is portrayed from left to right (and of course from bottom
to top).  We say that Figure \ref{den2} is a canonical representation 
of the dendrogram, implying that Figure \ref{den1} is not in canonical 
form.  In Figure \ref{den3}, the canonical representation has its 
non-terminal nodes labeled.  

\begin{figure}
\includegraphics[width=14cm]{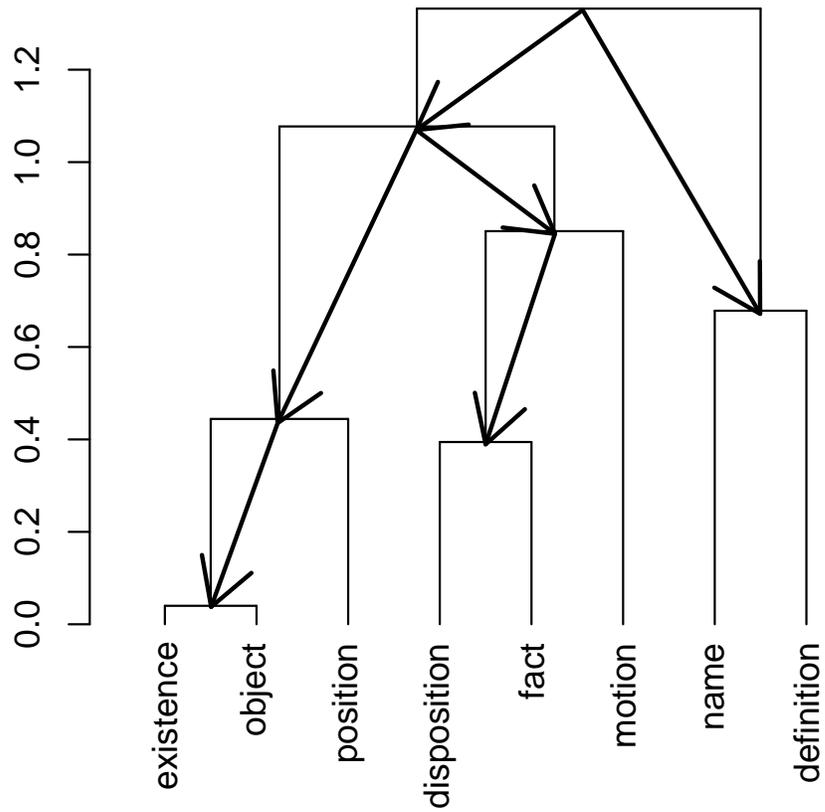}
\caption{Oriented binary tree is superimposed on the dendrogram.  The 
node at the arbitrary upper right location is not shown.  The oriented
binary tree defines an inorder or depth-first tree traversal.}
\label{den4}
\end{figure}

Next, Figure \ref{den4} shows a superimposed oriented binary rooted tree, 
on $n -1 $ nodes, which is isomorphic to the dendrogram on $n$ terminal 
nodes.  This oriented binary tree is an inorder traversal of the 
dendrogram.  
Sibson's [1973] ``packed representation'' of a dendrogram uses
just such an oriented binary rooted tree, in order to define a permutation
representation of the dendrogram.  
From our example, the packed representation
permutation can be read off as: $(13625748)$: for any terminal node indexed 
by $i$, with the exception of the rightmost which will always be $n$,
define $p(i)$ as the rank at which the terminal node is first
united with some terminal node to its right.
Discussion of combinatorial properties
of dendrograms, as related to such oriented binary rooted trees, and 
associated down-up and up-down permutations, can be found in Murtagh 
[1984a].  

\begin{figure}
\includegraphics[width=14cm]{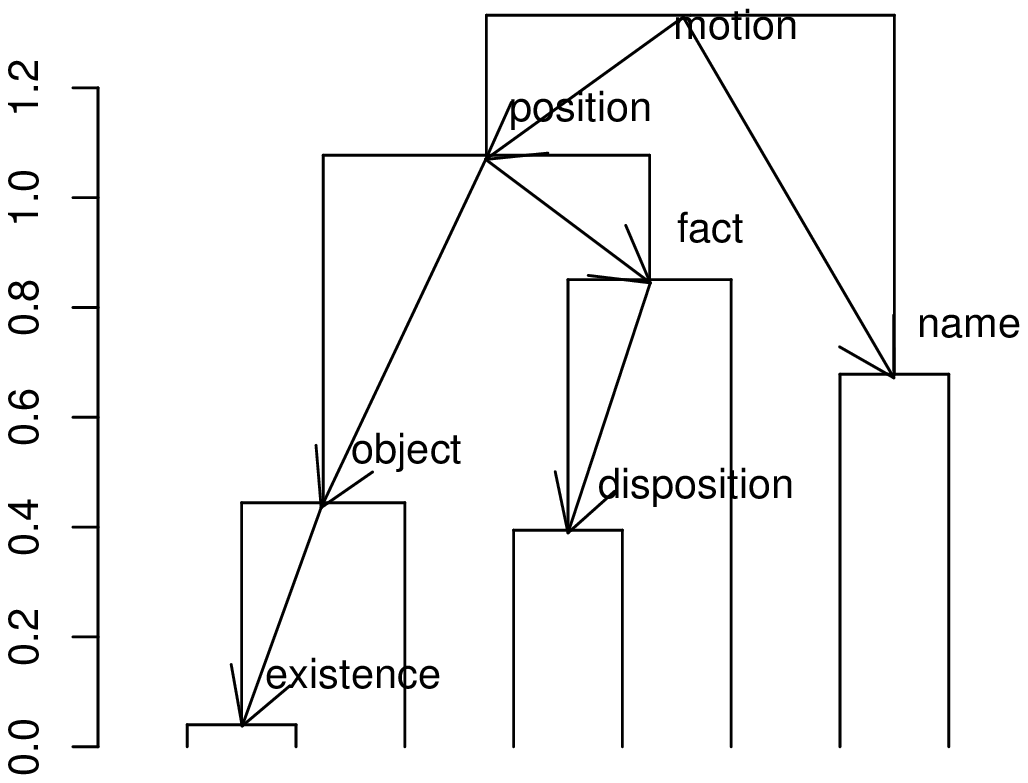}
\caption{The oriented binary tree with labeled nodes, using the procedure
discussed in the text.  The label ``definition'' is not shown and is 
to be located in the upper right, with an oriented tree arc directed at
``motion''.}
\label{den5}
\end{figure}

Finally, in Figure \ref{den5}, we ``promote'' terminal node labels to the 
nodes of the oriented tree.  We will use exactly the procedure used above
for defining a permutation representation of the oriented tree.  
First the left terminal label is promoted to its non-terminal node.  Next, 
the right terminal label is promoted as far up the tree as is necessary in 
order to find an unlabeled non-terminal node.  This procedure is carried 
out for all non-terminal labels, working in sequence from left to right 
(i.e., consistent with our canonical representation of the dendrogram).  
The rightmost label is not shown: it is at an arbitrary location in the 
upper right hand side, with a tree arc oriented towards the top non-terminal
node of the dendrogram, now labeled as ``motion''.

In this section, we have specified a consistent procedure for labeling 
the nodes of an oriented tree, starting from the labels associated with 
the terminal nodes of a dendrogram.  We start therefore with embedded 
clusters, and end up with terms and directed links between these terms.  
There is some non-uniqueness: any two labels associated with terminal nodes
that are left and right child nodes of one non-terminal node can be 
interchanged.  This clearly leads to a different label promotion outcome. 

Our promotion procedure was motivated by the permutation representation
of an oriented binary tree, as described above.  Here too we do not claim 
uniqueness of permutation representation.  But we do claim optimality 
in the sense of parsimony, and well-definedness.  

In the case of a multiway tree with very few distinct
levels, the promotion procedure becomes very simple, but continues to be
non-unique.  

\subsection{A New Approach to Deriving a Concept Hierarchy from a 
Dendrogram}
\label{sect52}

In the previous subsection, we discussed an algorithm which takes a 
hierarchical clustering, and hence a dendrogram, into an inorder tree
traversal, and hence a permutation of the set of terms used.  
The formal procedure discussed in the previous subsection suffers from
non-uniqueness: alternative permutations could be defined.  This leads us
to question the relationship of subsumption (or direction in the oriented 
tree).  In this section we will develop another approach which is even more
closely associated with the data that we are analyzing.  

We have already seen that triangle properties between triplets of 
points, or data objects, are fundamental to ultrametricity and hence 
to tree representation.  A dendrogram, representing a hierarchical clustering,
allows us to read off, for all triplets of points, 
either (i) isosceles triangles, with small base, or (ii) equilateral 
triangles, and (iii) no other triangle configuration.  The reason for the 
last condition is simply that non-isosceles, or isosceles with large base,
triangles are incompatible with the ultrametric, or tree, metric.  

We will leave aside for  the present the equilateral triangle case.  Firstly,
it implies that all 3 points are {\em ex aequo} in the same cluster.
Secondly, therefore we will treat them altogether as a concept cluster.
Thirdly, the equilateral case does not arise in the example we will now
explore.

\begin{figure}
\includegraphics[width=14cm,angle=270]{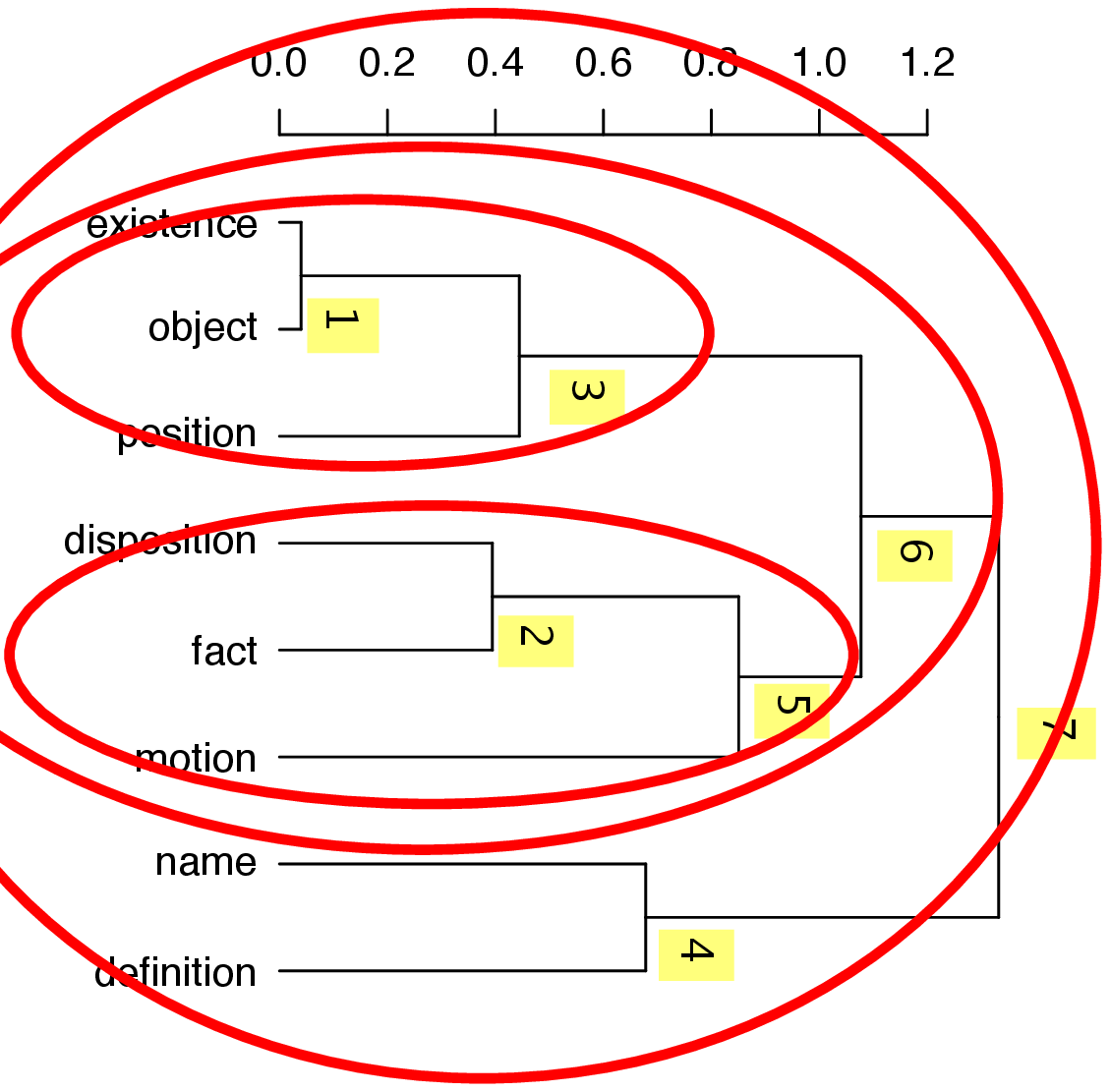}
\caption{Dendrogram as shown in Figures \ref{den2} and \ref{den3}, 
with clusters indicated by ellipses.  Shown here are ellipses covering 
the clusters at nodes 7, 6, 5, and 3.}
\label{den6}
\end{figure}

In Figure \ref{den6}, cluster number 3 indicates the following 
isosceles triangle with small base: ((existence, object) position).
Our notation is: ((x, y) z), such that triplet x, y, z has small base
x, y, and the side lengths x, z and y, z are equal.  This is necessarily
implied by relationships represented in Figure \ref{den6}.  So, 
motivated by this triangle view of the cluster number 3  part of the 
dendrogram we will promote ``position'' to the cluster number 3 node.  

Similarly we will promote ``motion'' to the cluster number 5 node.  

Note the consistency of our perspective on the cluster number 3 and 5 nodes
relative to how the associated terms here form an isosceles triangle with 
small base.

We will straight away generalize this definition.  
In any case of a node in the form of nodes 3 or 5, where we have a 
2-term left subtree, and a 1-term right subtree, where left and right 
are necessarily labeled in this way given the canonical representation
of the dendrogram, then: {\em the left subtree is dominated by the right 
subtree}.  

We will next look at cluster number 6 (remaining with Figure 
\ref{den6}).  As always for such trees, 
the node corresponding to this cluster has two subtrees, one to the left 
(here: 3) and one to the right (here: 5).  Since our dendrogram is in 
canonical form, any such node has a subtree with smallest non-terminal 
node level to the left; and the subtree which was more recently formed 
in the sequence of agglomerations to the right.  Based on either or 
 both of these criteria which serve to define what are the left and 
right subtrees we define the ordering relationship: 
{\em the left subtree is 
dominated by the right subtree}.  

\begin{figure}
FIGURE NOT AVAILABLE: SEE PDF VERSION OF PAPER AT 
www.cs.rhul.ac.uk/home/fionn/papers/auto\_onto.pdf
\caption{Concept relationships, ordered by dominance, derived from 
the dendrogram in Figure \ref{den6}.}
\label{concepts}
\end{figure}

Figure \ref{concepts} summarizes the concept relations that we can 
derive in a similar way from any dendrogram.  

\subsection{Demonstrator}

\begin{figure}
\includegraphics[width=14cm]{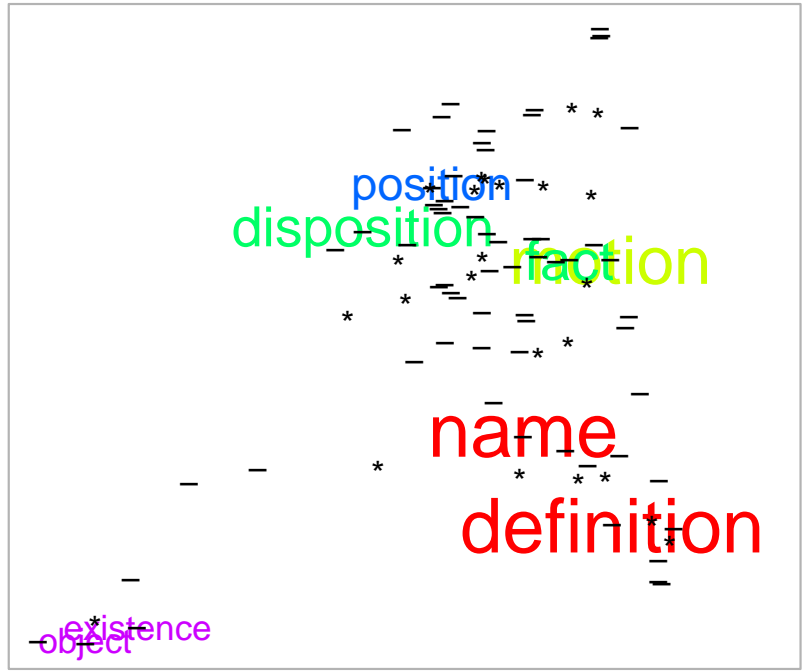}
\caption{The relationships displayed in Figure \ref{concepts}
are shown in decreasing size (and in rainbow colors, from red), with other
terms (in all, 66) displayed with a dash, and all text segments (in all, 24) 
represented by an asterisk.  The principal factor plane of a correspondence
analysis (based on the 24 text segments $\times$ 66 terms frequencies of 
occurence) output is used.}  
\label{aristontoex}
\end{figure}

Figure \ref{aristontoex} indicates how the concept relations, shown in 
Figure \ref{concepts}, are to be used.  

Firstly the term set is  summarized, using our 
selection of terms.  Scaling to large data sets is addressed in this way.

Secondly, in our interactive implementation (web address: \\
thames.cs.rhul.ac.uk/$\sim$dimitri/textmap), 
we allow the terms shown to continually move in a limited
way, to get around the occlusion problem, and we also allow magnification 
of the display area for this same reason.  

Thirdly, terms other than those
shown are highlighted when a cursor is passed over them.  

Next, double clicking 
on any term gives a ranked list of text segment names, 
ordered by frequency of occurrence by this term. 
Clicking on the text segment gives the actual 
text at the bottom of the display area.  

\subsection{Application with Ex Aequo Terms and Clusters}
\label{sect7}

We proceed now to a third  case study of this work, where we have a 
multiway hieararchy (and not a binary hierarchy) from the start.  We require a 
frequency of occurrence matrix which crosses the terms of interest with 
parts of a free text document.  For the latter we could well take 
documentary segments like paragraphs. 
O'Neill [2006] is a 660-word discussion of ubiquitous computing
from the perspective of human computing interaction.  With this short 
document we used individual lines (as proxies for the sequence of sentences)
as the component parts of the document.  There were 65 lines.  This 
facilitates retrieval of a small segment of such a single document. 
We chose this text to work with because it is a very small text (a 
single text compared to the data used in section \ref{sect6}, and 
a far smaller text compared to that used in section \ref{secorientree}).

Based on a list
of nouns and substantives furnished by the part-of-speech tagger (Schmid, 
1994), we focused on the following 30 nouns: 

{\tt support} = $\{$
``agents'',       ``algorithms'',   ``aspects'',      ``attempts'',     
``behaviours'',   ``concepts'',     ``criteria'',
``disciplines'', ``engineers'',    ``factors'',      ``goals'',        
``interactions'', ``kinds'',        ``meanings'',
``methods'',      ``models'',       ``notions'',      ``others'',       
``parts'',        ``people'',       ``perceptions'',
``perspectives'', ``principles'',   ``systems'',      ``techniques'',   
``terms'',        ``theories'',     ``tools'',   
``trusts'',       ``users'' $\}$.

This set of 30 terms was used to characterize through presence/absence 
the 65 successive lines of text, leading to correspondence analysis of 
the $ 65 \times 30$ presence/absence matrix.  This yielded then the 
definition of the 30 terms in a factor space.  In principle the rank of this
space (taking account of the trivial first factor in correspondence 
analysis, relating to the centering of the cloud of points) is min(
$65 - 1, 30 - 1$).  However, given the existence of
 zero-valued rows and/or columns, 
the actual rank was 25.  Therefore the full rank projection of the terms 
into the factor space gave rise to 25-dimensional vectors for each term, 
and these vectors are endowed with the Euclidean metric.  

Define this set of 30 terms as the {\em support} of the document.  
Based on their occurrences in the document, we generated the following
{\em reduced} version of the document, defined on this support, which 
consists of the following ordered set of 52 terms:

{\tt Reduced document} =
``goals''  ``techniques''   ``goals''  ``disciplines''  ``meanings''     
``terms''        ``others'' 
``systems''  ``attempts''     ``parts''  ``trusts''  ``trusts''    
``people''   ``concepts'' 
``agents''   ``notions''   ``systems''   ``people''  ``kinds''        
``behaviours''   ``people''  
``factors''      ``behaviours''   ``perspectives'' ``goals''        
``perspectives'' ``principles''   ``aspects''  
``engineers''    ``tools''        ``goals''        ``perspectives'' 
``methods''      ``techniques''   ``criteria'' 
``criteria''     ``perspectives'' ``methods''      ``techniques''   
``principles''   ``concepts''     ``models''   
``theories''     ``goals''        ``tools''        ``techniques''   
``systems''      ``interactions'' ``interactions''
``users''        ``perceptions''  ``algorithms''

This reduced document is just the ``time series'' of the nouns of interest
to us, as they are used in traversing the document from start to finish.
Each noun in the sequence of 52 nouns is represented by its
25-dimensional factor space vector.  

Out of 43 unique triplets, with self-distances removed, we found 31 to 
respect the ultrametric 
inequality, i.e.\ 72\%.  Our measure of ultrametricity 
of this document, based on the support used, was thus 0.72.  

For a concept hierarchy we need an overall fit to the data.  Using 
the Euclidean space perspective on the data, furnished by correspondence
analysis, we can easily define a terms $\times$ terms distance matrix; 
and then hierarchically cluster that.  Consistent with our analysis we
recode all these distances, using the CvNC mapping onto $\{ 1, 2 \}$ for 
unique pairs of terms.  

%Note that this is tantamount to having a window,
%$r$ (in the notation used in section \ref{sectext}), encompassing all of the
%reduced document.  It is also interesting to check the ultrametricity 
%coefficient here.  This means therefore the ultrametricity coefficient 
%in the window length $n$ case, versus the ultrametricity coefficient 
%in the window length 3 case.  The latter was seen to be (from exhaustive 
%calculation) above, 0.72.  For the window length $n$ case, we sampled
%2000 triplets, and found the ultrametricity coefficient to be 0.56.  
%Since the linear order is of greater ultrametric (hence, hierarchical) 
%structure, an evident question arises as to whether it should be used 
%as the basis for a retrieved overall or global hierarchy.  We do not do 
%this, however, because the greater hierarchical structure comes as the cost
%of being overly fragmentary.  

Now approximating a global ultrametric from below, achieved 
by the 
single linkage agglomerative hierarchical clustering method (and this 
best fit from below, termed the subdominant or maximal inferior 
ultrametric, is optimal), and an 
approximation from above, achieved by the complete linkage agglomerative 
hierarchical clustering method (and this best fit from above, termed a 
minimal superior ultrametric, is 
non-unique and hence is one of a number of best fits from above), will be
identical if the data is fully ultrametric-embeddable.  If we had an
ultrametricity coefficient equal to 1 -- we found it to be 0.72 for this
data -- then it would not matter what agglomerative hierarchical 
clustering algorithm (among the usual Lance-Williams methods) that we 
select.  

\begin{figure}
\centering
\includegraphics[width=14cm,angle=270]{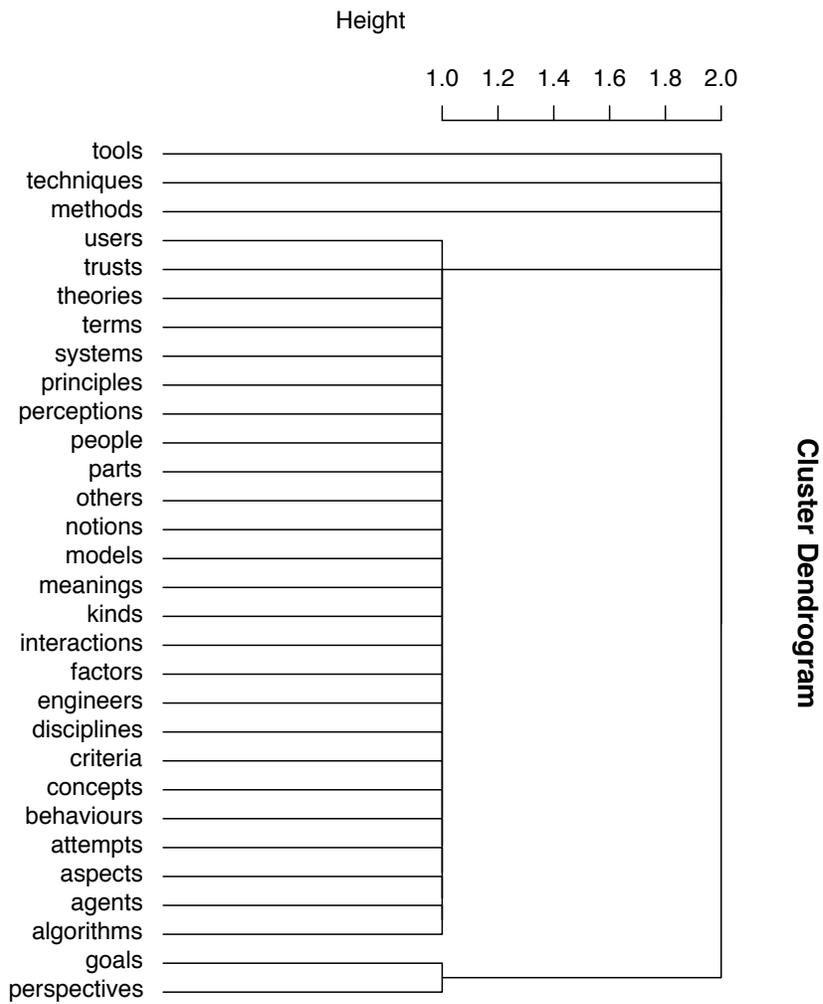}
\caption{Single (or identically, complete) linkage hierarchy of 30 
terms, comprising the support of the document, based on (i) 
``no change/change'', or CvNC, metric recoded (ii) 25-dimensional Euclidean  
representation.}
\label{fig1}
\end{figure}

In fact, we found, with an ultrametricity coefficient equal to 0.72, that 
the single and complete linkage methods gave an identical result.  This 
result is shown in Figure \ref{fig1}.  

A convenient label promotion procedure to apply here is first to 
re-represent the terminal labels from left to right as: 
$\{$ ``users'', ``trusts'', $\dots$, ``agents'', ``algorithms'' $\}$; 
$\{$ ``goals'', ``perspectives'' $\}$; $\{$ ``tools'' $\}$; 
$\{$ ``techniques'' $\}$; and $\{$ ``methods'' $\}$.  This is the canonical 
form, with ordering of left and right subtrees now extended to all 
subtrees.

Next, we must in fairness take  the nodes at level 2 as being {\em ex aequo}, 
$\{$ ``tools'' $\}$;
$\{$ ``techniques'' $\}$; and $\{$ ``methods'' $\}$.  Similarly at level 
1, we also have two clusters that are {\em ex aequo}: 
$\{$ ``users'', ``trusts'', $\dots$, ``agents'', ``algorithms'' $\}$; and
$\{$ ``goals'', ``perspectives'' $\}$.

\begin{figure}
\centering
FIGURE NOT AVAILABLE: SEE PDF VERSION OF PAPER AT 
www.cs.rhul.ac.uk/home/fionn/papers/auto\_onto.pdf
\caption{Concept hierarchy derived, using our methodology, from 
Figure \ref{fig1}.  See accompanying discussion for details.}
\label{ubiq}
\end{figure}

Figure \ref{ubiq} shows our resulting scheme where level 1 clusters 
dominate level 2 clusters.  

This provides our ontology.  The granularity of this one document 
is, as mentioned above, line-based, and there are 65 lines in all.  
Hence retrieval of one or more of these document snippets is supported, and
the ontology is based on a 30-noun document support.  

\section{Conclusions}

Having first appraised text collections in terms of their local hierarchical
structure, we then proceeded in this work 
to show how this new methodology could be employed for a
wide range of tasks that include: 

\begin{itemize}
\item finding salient pairs and triplets of terms, which are
not necessarily in sequence; 
\item permitting us to consider any given text 
as a whole with all pairwise relationships between terms, or alternatively
as a time series with relationships restricted to terms that are 
successive in sequence; 
\item passing seamlessly from the exploration of 
local hierarchical structure to global hierarchical structure; 
\item especially 
when global hierarchical structure is manifest, being able to use any 
of a wide range of agglomerative clustering criteria to furnish the same
resultant hierarchy; 
\item determining a hierarchy of concepts from the embedded, partially ordered
subsets provided by a hierarchical clustering; 
\item obtaining unique results when given a 2-way hierarchical clustering
tree, and then readily generalizing this to the practical case of
multiway trees; 
\item exemplifying an efficient and an effective textual data processing
pipeline; and 
\item through the measurement of local hierarchical structure, having 
available an approach to validating the appropriateness of any data 
for this data analysis pipeline.  
\end{itemize}

By analysis of text through local hierarchical relationships between 
terms we determine extensive internal textual structure, 
without being stifled by the more traditional approach of fitting 
some global structure, such as a hierarchy, to the text.  A (local) 
hierarchical structure is a powerful one: it includes peer as well 
as subsumption types of relationships.  

We stress that we can find very pronounced hierarchical structures of 
this sort if we encode the text is novel ways.  An example is to start 
with a Euclidean spatial embedding of the terms and documents (or 
segments of a document), which is quite traditional; and then look at 
interrelationships between terms using ``relatively close/similar'' 
versus ``relatively distant/new'' (and this alone can be shown to have 
metric properties).  Another example of an encoding-related strategy 
is not to take into consideration all interrelationships between terms, 
but only between successive terms, and thereby view the text as a 
particular type of time series. 
User interactivity with the system is to select the terms of interest 
(people's personal names, industrial product names, location or venue 
names, etc.).  The interrelationships between these terms are then 
explored through their local hierarchical links.  

Our general application targeted is, as stated in Murtagh et al.\ [2003], 
to have readily available a self-description of data, as a basis for 
visually-based interactive and responsive querying of, retrieval from,  
and navigation of data collections.  

%In the context of the WS-Talk project,
%within which this work was carried out, we have developed an interactive
%visualization-based data description and information retrieval tool called
%NodeMap [Contreras et al.\ 2006].  The work presented in this article 
%supports the building of ontologies and hence 
%the operation and use of this tool.

\subsection*{Acknowledgement}

This work was carried out in the context of the European Union Sixth 
Framework project, ``WS-Talk, Web services communicating in the language
of their community'', 2004--2006.  Pedro Contreras and Dimitri Zervas
contributed to this work.  The Textmap demonstrator was developed by 
Dimitri Zervas.

\section*{References}
\begin{description}

\item[] ABOU ASSALI, A. and ZANGHI, H. 2006.  
Automated metadata hierarchy derivation.  In 
{\em  Proc. IEEE ICTTA06} (Damascus, Syria).

\item[] AHMAD, K. and GILLAM, L. (2005).  
Automatic ontology extraction from 
unstructured texts.  In {\em ODBASE 2005}.

\item[] 
BENZ\'ECRI, J.P. (1979). {\em L'Analyse des Donn\'ees, Tome I Taxinomie,
Tome II Correspondances}, 2nd ed.\ (Dunod, Paris).

\item[] CHUANG SHUI-LUNG and CHIEN LEE-FENG (2005).
Taxonomy generation for text
segments: a practical web-based approach, ACM Transactions on Information 
Systems. 23, 363--396.

%\item[] CONTRERAS, P., ZERVAS, D. and MURTAGH, F. (2006). 
%NodeMap ontology visualization and information retrieval tool, \\
% http://thames.cs.rhul.ac.uk/$\sim$wstalk/prototype.html

\item[] DE SOETE, G. (1986). 
A least squares algorithm for fitting an ultrametric tree
to a dissimilarity matrix.
{\em Pattern Recognition Letters}, 2, 133--137.

\item[] DENNY, M. (2004). Ontology tools survey, Revisited, at  \\ 
http://www.xml.com/pub/a/2004/07/14/onto.html, July 2004.  

\item[] DOYLE, L.B. (1961). Semantic road maps for literature searches.
{\em Journal of the ACM}. 8, 553--578.

\item[] GANESAN, P., GARCIA-MOLINA, H. and WIDOM, J. (2003). 
Exploiting hierarchical
domain structure to compute similarity.  {\em ACM Transactions on 
Information Systems}.  21, 64--93.

\item[] G\'OMEZ-P\'EREZ, A., FERN\'ANDEZ-L\'OPEZ, M. and CORCHO, O.
(2004).  
{\em Ontological Engineering (with Examples from the Areas of Knowledge
Management, e-Commerce and the Semantic Web)} (Springer, Berlin). 

\item[] GRUBER, T. (2001).  What is an ontology?, 
http://www-ksl.stanford.edu/kst/what-is-an-ontology.html, Sept. 2001.

\item[] JANOWITZ, M.F. (2005).  Cluster analysis based on abstract posets, 
preprint.  Also presentation, ENST de Bretagne, 30 Oct. 2004; DIMACS, 9 
Mar. 2005; and SFC, 31 May 2005.  

\item[] MAEDCHE, A. (2006).  {\em Ontology Learning for the Semantic Web}
(Kluwer, Dordrecht).

\item[] MURTAGH, F. (1984a).  Counting dendrograms: a survey.
{\em  Discrete Applied Mathematics}, 7, 191--199.

\item[] MURTAGH, F. (1984b). 
Structures of hierarchic clusterings: 
implications for information retrieval and for multivariate 
data analysis.  {\em Information Processing and Management}, 20, 
611--617.

\item[] MURTAGH, F. (2004). 
On ultrametricity, data coding, and computation,
{\em Journal of Classification}, 21, 167--184.

\item[] MURTAGH, F. (2005a). Identifying the ultrametricity of time series,
{\em European Physical Journal B}, 43, 573--579.

\item[] MURTAGH, F. (2005b).  {\em Correspondence Analysis and Data 
Coding with R and Java} (Chapman \& Hall/CRC).

\item[] 
MURTAGH, F. (2006a). Ultrametricity in data: identifying and exploiting 
local and global hierarchical structure, arXiv:math.ST/0605555v1 19 May 2006.

\item[] MURTAGH, F. (2006b).  The Haar wavelet transform of a dendrogram, \\
arXiv:cs.IR/0608107v1 28 Aug 2006, forthcoming in 
{\em Journal of Classification}.

\item[] MURTAGH, F. (2006c).  Symbolic dynamics in 
text: application to automated 
construction of concept hierarchies, {\em Festschrift for Edwin Diday}, 
in press.  Available at:  www.cs.rhul.ac.uk/home/fionn/papers

\item[] MURTAGH. F. (2006d).  Visual user interfaces, interactive maps, 
references,
theses, http://astro.u-strasbg.fr/$\sim$fmurtagh/inform

\item[] MURTAGH, F., TASKAYA, T., CONTRERAS, P., MOTHE, J. and ENGLMEIER, K.
(2003). 
Interactive visual user interfaces: a survey, {\em Artificial Intelligence 
Review}, 19, 263--283.

\item[] O'NEILL, E. (2006).  Understanding ubiquitous computing: a view from
HCI, in Discussion following R. Milner, Ubiquitous computing:
how will we understand it?, {\em Computer Journal}, 49, 390--399.

\item[] SCHMID, H. (1994).  Probabilistic part-of-speech tagging using decision
trees.  In {\em Proc. Intl. Conf. New Methods in Language Processing} 
(see TreeTagger site).

\item[] TreeTagger, www.ims.uni-stuttgart.de/projekte/corplex/TreeTagger/
DecisionTreeTagger.html

\item[] SIBSON, R. (1973).  SLINK: an optimally efficient 
algorithm for the single-link
cluster method, {\em Computer Journal}, 16, 30--34.

\item[] WACHE, H., V\"OGELE, T., VISSER, U., STUCKENSCHMIDT, H., SCHUSTER,
G., NEUMANN, H. and H\"UBNER, S. (2001).  
Ontology-based integration of information -- 
a survey of existing approaches.  In {\em 
Proceedings of the IJCAI-01 Workshop: Ontologies and Information Sharing}.

\end{description}
\end{document}